\documentstyle[11pt]{article}
\newlength{\bredde}
\def\slash#1{\settowidth{\bredde}{$#1$}\ifmmode\,\raisebox{.15ex}{/}
\hspace*{-\bredde} #1\else$\,\raisebox{.15ex}{/}\hspace*{-\bredde} #1$\fi}

\newcommand{\beq}{\begin{equation}}
\newcommand{\eeq}{\end{equation}}
\newcommand\beqn{\begin{eqnarray}}
\newcommand\eeqn{\end{eqnarray}}
\newcommand{\nn}{\nonumber}
\newcommand{\noi}{\vspace{12pt}\noindent}
\newcommand{\Tr}{\mbox{Tr}}
\newcommand{\de}{\delta}
\newcommand{\th}{\theta}

\newcommand{\Pd}{{\cal P}_\de}

\newcommand{\Pc}{{\cal P}}
\newcommand{\Pl}{{\cal P}_l}
\newcommand{\Zd}{{\cal Z}_\de}

\newcommand{\Z}{{\cal Z}}
\newcommand{\Zl}{{\cal Z}_l}
\newcommand{\dM}{{\cal D}M}
\newcommand{\rd}{\rho_\de}

\newcommand{\rl}{\rho_l}
\newcommand{\Pint}{{\mathbf{-}}\!\!\!\!\!\!\int}

\def\l{\label}
\def\r{\ref}
\def\la{\lambda}
\def\al{\alpha}

\begin{document}
\topmargin -1.4cm
\oddsidemargin -0.8cm
\evensidemargin -0.8cm
\title{\Large{{\bf Compact support probability distributions in 
random matrix theory}}}

\vspace{2.5cm}

\author{~\\{\sc G. Akemann $^{(1)}$, G.M. Cicuta $^{(2)}$, L. Molinari $^{(3)}$
 and G. Vernizzi $^{(2)}$ }\\~\\
 (1) Centre de Physique Th\'eorique CNRS\\
Case 907 Campus de Luminy\\F-13288 Marseille Cedex 9, France\\~\\
(2) Dipartimento di Fisica\\Viale delle Scienze, I-43100 Parma, Italy\\
and INFN Gruppo collegato di Parma\\~\\
(3) Dipartimento di Fisica and INFN\\Via Celoria 16, I-20133 Milano, Italy}
\maketitle
\vfill
\begin{abstract}
We consider a generalization of the fixed and bounded trace ensembles
introduced by Bronk and Rosenzweig up to an arbitrary polynomial potential.
In the large-$n$ limit we prove that the two are equivalent and that their
eigenvalue distribution coincides with that of the "canonical" ensemble
with measure exp[-$n$Tr V(M)]. The mapping of the corresponding phase 
boundaries is illuminated in an explicit example. In the case of a 
Gaussian potential we are able to derive exact expressions for the
one- and two-point correlator for finite $n$, having finite
support. 
\end{abstract}
\vfill

\begin{flushleft}
CPT-98/P.3675\\
cond-mat/9809270
\end{flushleft}
\thispagestyle{empty}
\newpage
\setcounter{page}{1}

\setcounter{equation}{0}
\section{Introduction}

Random matrix ensembles have been extensively studied since the
early works of Wigner and Dyson, as effective mathematical
reference models for the description of statistical properties in the
spectra of complex physical systems, ranging from such diverse
areas as nuclear resonances or quantum billiards to mesoscopic
transport or quenched QCD. Even a cursory glance at some recent
review monographs   \cite{ambjorn}, \cite{guhr},  \cite{beenakker},
 \cite{zinn}, shows the impressive development of analytical tools
and the variety of applications to physical systems reached in
the past decade and a combined bibliography, although very
incomplete, of over a thousand papers.\\ Historically the matrices
of the ensemble belong to one of three classes, they are real
symmetric or complex-Hermitian or with quaternionic entries but
in recent years other ensembles , like complex non-Hermitian or
real non-symmetric matrices have been studied. To keep our
paper as simple as possible, we restrict ourselves to complex-Hermitian
matrices, although the results of this paper apply also to the
other two traditional ensembles with minimal changes.\\
A random matrix ensemble is defined by the joint probability
density for the independent entries of the matrix. In a large number
of papers, particularly those related to two-dimensional quantum
gravity, the probability density has the form
\beqn
{\cal P}(M) &\equiv& \frac{1}{{\cal Z}}\,  e^{-\beta \Tr \, V( M)}
\label{i.1}
\eeqn
where $V(x)$ is a polynomial. Since this  probability density is invariant
under the similarity transformation $M= U \Lambda U^{\dag} $ which
diagonalizes the matrix $M$, most problems 
 may be formulated in terms of
the joint probability density for the eigenvalues 
\beqn
{\cal P}(\lambda_1,..,\lambda_n)&\equiv& \frac{1}{z}\, 
\Delta^2_n(\la)
\,  e^{-\beta \sum_1^n \, V(\lambda_i )}
\label{i.2}
\eeqn
and may be called eigenvalue models. We shall call the
probability density (\ref{i.1}) or (\ref{i.2}) the "canonical"
density. In the analysis of the "large-$n$ " limit
of observables, evaluated with the "canonical" probability density,
the method of orthogonal polynomials  \cite{bessis} proved to be most
effective.  In the present paper we study matrix ensembles
defined by the probability density
\beqn
\Pd(M) &\equiv& \frac{1}{\Zd}\, \delta \bigg( A^2-\frac{1}{n}{ \Tr}\, V( M) 
\bigg)
\label{i.3}
\eeqn
and the closely related probability density where the step function
replaces the Dirac delta function. We follow the classic book by
Mehta   \cite{mehta} 
and call collectively these
models {\bf (generalized) restricted trace ensembles}.
They are a generalization of
ensembles studied long ago by Rosenzweig and Bronk
 \cite{rosenzweig} where only the case $V(x)=x^2$ was considered.
While the ensemble is still invariant under the unitary transformation
which diagonalizes the random matrix, the  method of orthogonal polynomials
cannot be directly applied because the constraint of the delta function
introduces an additional interaction among the eigenvalues.
Restricted trace ensembles seem to us interesting for several
features : the interaction among eigenvalues is introduced
through a constraint very similar to the non linear sigma model
in quantum field theory, the spectral density has compact support
both for finite $n$ and in the "large-$n$ " limit (unlike the
usual Gaussian random model), and they relate to "canonical"
probability densities (\ref{i.1}) or (\ref{i.2}) just in the same
way as the microcanonical ensemble is related to the
canonical ensemble in statistical mechanics.\\
The effectiveness of random matrix theory is related mainly to
"universal" properties of some observables, that is independence,
in the "large-$n$ " limit, of some observables from the chosen
probability density.  The ensemble averaged  density of eigenvalues
$\rho(\lambda)=\frac{1}{n} \Tr <\delta(\lambda-M)>$ , 
in "canonical"
eigenvalue models (\ref{i.1}) or (\ref{i.2}) is known to depend from
the chosen function $V(x)$,  yet a number of critical exponents
deduced from the spectral density were shown to be independent from 
the details of the chosen function $V(x)$. Much earlier
the density-density connected
correlator $\rho_c (\lambda, \mu)$
\beqn
\rho_c (\lambda, \mu) &\equiv& <  \,{1\over n}\mbox{ Tr}\delta 
(\lambda -M)\,
{1\over n}\mbox{ Tr}\delta (\mu -M) \,> -
<  \,{1\over n}\mbox{ Tr}\delta (\lambda -M)\,>
\,<  \,{1\over n}\mbox{ Tr}\delta (\mu -M) \,> \, =
\nonumber \\
&=&\rho (\lambda, \mu)
-\rho (\lambda) \,\rho (\mu)
\label{i.4}
\eeqn
was shown to have "local universality" properties , that is for
$|\lambda-\mu | \sim O(1/n) $ and far from the extrema of the
support of  the spectral density,
in the "large-$n$" limit.  This was the basis for the use of
random matrix theory for statistical fluctuations of observables
around their mean values.
 Other forms of universality were derived more recently by several
authors, including a form of "wide correlator"  which depends
on the "canonical" potential $V(x)$ only through the extrema
of  the spectral density. The proof by Beenakker and a list
of other authors is recalled in sect.1 D of ref. \cite{beenakker}.\\
In sect.2 we exploit a scale transformation, already used by Rosenzweig in
a more limited extent,
to relate observables in restricted trace ensembles where $V(x)=x^2$
 with the corresponding
ones in the random Gaussian model.  This allows explicit evaluations for
 the spectral
density and the two point correlators  for finite $n$. \\
We then consider a generalization of the restricted trace ensembles 
to a generic $V(x)$ in sect.3.
There a  very general proof of the equivalence, in the large-$n$ 
limit, of the generalized restricted trace ensembles with the corresponding
"canonical" ones is presented.  This proof is a wide generalization
of the old result of the  equivalence, in the large-$n$ limit, of
the restricted trace ensembles with the random Gaussian model .\\
Unlike the original  restricted trace ensembles, the 
generalized ensembles have a non trivial
phase diagram in the  large-$n$ limit.  Despite the equivalence
shown in sect.3 with "canonical"  probability distributions, the mapping of 
parameters in "equivalent" models is one-to-one only in the
"perturbative phase". We show this 
in detail in one example of phase diagram in sect.4.\\
Let us stress that the present
paper is concerned with derivation of exact analytic results for
the probability  distributions we consider. Applications of physical
interest are deferred to a future paper.
While this paper was being written, we were informed of a poster 
presented by T.Nagao  at StatPhys 20, discussing generalized
fixed trace ensembles of random matrices. There the old model
by Rosenzweig is generalized by considering  a joint probability
density of eigenvalues of the form
\beqn
{\cal P}(\lambda_1,..,\lambda_n)&\equiv& \frac{1}{z} \, 
\Delta^{\beta}_n(\la) \; \prod_1^n  \lambda^{\alpha}_i
\, \delta( A^2- \sum_1^n \, \lambda^2_i ) \ \ .
\label{i.5}
\eeqn
This study has very little overlap with the present paper.
\vspace{1.5cm}

\setcounter{equation}{0}
\section{Restricted trace ensembles  at finite n.}

Let us begin by describing the most relevant features of two closely
related ensembles : the fixed trace and the bounded trace 
 ensembles. Let $M$ be a $n$ x $n$ Hermitian matrix.
The {\bf fixed trace ensemble} corresponding to the Gaussian
model is defined by the probability
distribution
\beqn
\Pd(M) &\equiv& \frac{1}{\Zd}\, \delta(A^2-\frac{1}{n}{ \Tr}\, M^2) \quad ,
\nn \\
\Zd &\equiv & \int {\cal D}M \, \delta(A^2-\frac{1}{n}{ \Tr}\, M^2)
=\left({\frac{1}{2}}\right)^{(n^2-n)/2}  
\;\omega_{n^2}\;
\frac{ (A \sqrt{n})^{n^2} }{2 A^2} \ \ ,
\label{c.1}
\eeqn
where $ {\cal D}M \equiv \prod_{i=1,..n}dM_{ii}\,  \prod_{i>j} \mbox{ Re} 
\,dM_{ij}\,
\mbox{ Im} \,dM_{ij} $ , $\ \omega_{n^2} = \frac{2 \, \pi^{n^2/2} }
{ \Gamma (\frac{n^2}{2})} 
 $ is the surface area of the unit sphere
in $n^2$ dimensions, 
and the factor $\frac{1}{n}$ has been introduced in view of the
large-$n$ limit.\\
Expectation values of $O(n^2)$ invariant amplitudes are trivially
evaluated for every $n$ as for instance
\beq
<({ \Tr}\,M^2)^k\,>_{\de }\ \equiv \int {\cal D}M\; ({ \Tr}\,M^2)^k\,\;\Pd(M)
=(n A^2)^k \ \ .
\label{c.2}
\eeq
However we are interested in more general expectation values, which are
functions of the distribution of eigenvalues. They may be evaluated from
the joint probability distribution $\Pd(\la_1,..,\la_n)$
which is obtained from eq.(\ref{c.1}) after integration of the unitary 
degrees of freedom
\beqn
\Pd(\la_1,..,\la_n) &=&
\frac{1}{z_{\de }}\,  \Delta^2_n (\lambda) \, \delta(
A^2-\frac{1}{n}\sum_{i=1}^{n} \,\la_i^2)
\; , \quad 
\Delta_n(\lambda) \, \equiv \prod_{1 \leq r<s \leq n} (\la_r-\la_s) =
\det \big[\la_i^{j-1} \big] \; ,
\nn \\
z_{\de }  &\equiv & \int_{-\infty}^{\infty}\!
\prod_{i=1}^n d\la_i\;\Delta^2_n (\lambda) \, \delta(
A^2-\frac{1}{n}\sum_{i=1}^{n} \,\la_i^2)=
\bigg(\frac{A^2}{2}\bigg)^{ \frac{n^2}{2} -1} \,  \frac{n^{n^2/2}}{2}
\frac{(2 \pi)^{n/2} }{\Gamma(\frac{n^2}{2})}
\; \prod_{j=1}^n \, j!\  \ .
\nn \\
\label{c.4}
\eeqn
Closely related to this matrix ensemble is the {\bf bounded trace ensemble}. It
 is defined by the probability
distribution
\beqn
{\cal P}_{\theta}(M) &\equiv & \frac{1}{{\cal Z}_{\theta} }\,
 \th(A^2-\frac{1}{n}{ \Tr}\, M^2) \quad ,
\nn \\
{\cal Z}_{\theta} &\equiv & \int {\cal D}M \, \th(A^2-\frac{1}{n}{\Tr}\, M^2)=
\left({\frac{1}{2}}\right)^{(n^2-n)/2}  
\;\omega_{n^2}\;
\frac{ (A \sqrt{n})^{n^2} }{n^2}\ \ .
\label{c.5}
\eeqn
In the same way of eq.(\ref{c.2}), one easily finds
\beqn
<({ \Tr}\,M^2)^k\,>_{\th }\ =\bigg( \frac{n^{2+k}}{n^2+2 k} \bigg) \, A^{2 k} 
\ \ .\label{c.6}
\eeqn
which exhibits the usual factorization of $O(n^2)$ invariant amplitudes only
in the large-$n$ limit.
In order to evaluate expectations which only depend on the distribution 
probability of the
eigenvalues, one may use the joint probability distribution, analogous to 
eq.(\ref{c.4}) :
\beqn
{\cal P}_{\theta}(\la_1,..,\la_n)&=&
\frac{1}{z_{\th }}\,   \Delta^2_n (\lambda)\, \th(
A^2-\frac{1}{n}\sum_1^{n} \,\la_i^2) \ \ , 
\nn \\
z_{\th }  &\equiv & \int_{-\infty}^{\infty}\!\prod_{i=1}^n
d\la_i\;  \Delta^2_n (\lambda)\, \th(
A^2-\frac{1}{n}\sum_1^{n} \,\la_i^2) =
\bigg(\frac{A^2}{2}\bigg)^{ \frac{n^2}{2} } \,  n^{n^2/2}
\frac{(2 \pi)^{n/2} }{\Gamma(\frac{n^2}{2} +1 )}
\; \prod_{j=1}^n  \,j! \ \ .
\nn \\
\label{c.7}
\eeqn
Of course the two ensembles are related by a simple differential equation. 
Since
\beqn
\frac{\partial}{\partial A^2}{\cal P}_{\theta}(M)= 
\frac{\Zd}{{\cal Z}_{\theta}}
\bigg[ \Pd(M)-{\cal P}_{\theta}(M)\bigg]=
\frac{n^2}{2 A^2}  \bigg[ \Pd(M)-{\cal P}_{\theta}(M)\bigg]
\nn
\eeqn
one easily obtains a simple relation between the two expectations for
any generic observable
\beqn
<O(M)>_{\de }\ = \bigg( 1+\frac{2 A^2}{n^2} \frac{\partial}{\partial A^2}
 \bigg)
<O(M)>_{\th } \ \ .
\label{c.8}
\eeqn
A remarkable feature of both the "fixed trace ensemble"  and the "bounded 
trace ensemble" is that the
density of states $\rho(\la)$ has compact support for any
$n$ , finite or infinite. 
We here obtain the exact  expression of the eigenvalue distribution of the
fixed trace ensemble for any value of $n$, based on the known results
for the Gaussian model.\\
Let us first recall a few useful formulas of the Gaussian model. The
partition function and the eigenvalue density are
\beqn
 z_G = \int d\lambda_1\ldots d\lambda_n  \,  \Delta^2_n   \,
e^{-a(\lambda_1^2+\ldots + 
\lambda_n^2)}  \ \ ,
\label{c.9}
\eeqn
\beqn
 \rho_G (\lambda ) =e^{-a\lambda^2}{1\over z_G}\int d\lambda_1\ldots 
d\lambda_{n-1} \Delta^2_{n-1} e^{-a(\lambda_1^2+\ldots +\lambda_{n-1}^2)}
\prod _{i=1}^{n-1} (\lambda-\lambda_i)^2 \ \ ,
\label{c.10}
\eeqn
where the positive parameter $a$ is arbitrary, and for shortness we set
$\Delta^2_n\equiv \Delta^2(\lambda_1,\ldots , \lambda_n)$. Both integrals
may be computed for finite $n$ by means of orthogonal polynomials, which
in this case are the Hermite ones:
\beqn
z_G = {{(2\pi)^{n/2}}\over {(2a)^{n^2/2}}}\prod_{k=1}^n k! \  \ , \quad
\quad  \rho_G(\lambda )= \sqrt {a\over \pi}  e^{-a\lambda^2}
{1\over n}\sum_{k=0}^{n-1} {{H_k^2(\lambda\sqrt a)}\over {2^k k!} }\ \ .
\label{c.11}
\eeqn
Instead of evaluating the sum by means of the Christoffel-Darboux formula, it
is useful for our discussion to use the expansion
\beqn
[ H_k(x)]^2 = \sum_{\ell=0}^k {{(k!)^2 2^{k-\ell} }
\over {(\ell !)^2 (k-\ell)!}} H_{2\ell}(x)
\label{c.12}
\eeqn
to obtain, with some simple algebra:
\beqn
\rho_G(\lambda ) =
\sqrt {a\over \pi} e^{-a\lambda^2}{1\over n}\sum_{k=0}^{n-1}
\pmatrix {n\cr k+1\cr } {{H_{2k}(\lambda\sqrt a)}\over {2^k k!}} \ \ .
\label{c.13}
\eeqn
To study the integrals for the fixed trace ensemble, it is convenient to
adopt the following notation. Let us denote by $\omega (n,R)$ the
surface of the sphere in $R^n$ of radius $R$, and by $da_n$ the element of
surface integration. The partition function and the eigenvalue density,
for $|\lambda |\le R$, are:
\beqn
z_{\de } = \int_{-\infty}^{\infty}\!
\prod_{i=1}^n d\la_i\; \Delta^2_n (\lambda)\, \delta(
A^2-\frac{1}{n}\sum_{i=1}^{n} \,\la_i^2)= \frac{n}{2 R}
\int_{\omega (n,R)} da_n \,  \Delta^2_n
\quad , \quad R^2 \equiv n A^2\ \ ,
\label{c.14}
\eeqn
\beqn
\rho_{\delta } (\lambda ) &=&  \frac{1}{  z_{\de }} \int \prod_{i=1}^n d\la_i\;
  \Delta^2_n \, \delta (\lambda-\lambda_n)\, \delta (A^2-\frac{1}{n}
\sum_{i=1}^{n} \,\la_i^2) =
\nonumber \\
 &=& \frac{n}{2  z_{\de } \sqrt{R^2-\lambda^2}}
\int_{\omega (n-1,\sqrt {R^2-\lambda^2})}da_{n-1}
\Delta^2_{n-1}\prod_{i=1}^{n-1}(
\lambda-\lambda_i)^2 \ \ .
\label{c.15}
\eeqn
After a change of scale, to restrict  both integrals to the surface of unit 
radius:
\beqn
z_{\de } = \frac{n}{2} R^{n^2-2}\int_{\omega (n,1)} da_n \, \Delta^2_n \ \ ,
\label{c.16}
\eeqn
\beqn
\rho_{\delta }(\lambda)= \frac{n}{2 z_{\de }} (R^2-\lambda^2)^{ \frac{1}{2}
(n^2-3)}
\int_{\omega (n-1,1)}da_{n-1}  \Delta^2_{n-1} \prod_{i=1}^{n-1}(
{\lambda\over { \sqrt {R^2-\lambda^2} }}-\lambda_i)^2 \ \ .
\label{c.17}
\eeqn
Let us first evaluate the partition function. We start from the integral
expression (\ref{c.9})
 for $z_G$, and change to spherical variables with radial
component $r$. The volume element is $r^{n-1}dr da_n$, and
$\Delta^2(\lambda_1, \ldots , \lambda_n) = r^{n(n-1)}\Delta^2(\lambda_1/r,
\ldots ,\lambda_n/r)$. Therefore we have:
\beqn
 z_G = \int_0^\infty dr r^{n^2-1} e^{-ar^2} \int_{\omega (n,1)} da_n
\Delta^2_n \ \ .
\label{c.18}
\eeqn
The surface integral is the same appearing in (\ref{c.16} ), and we conclude:
\beqn
z_{\de } = z_G \, 
\frac{ n ( R \sqrt{a})^{n^2} }{R^2 \, \Gamma(\frac{n^2}{2}) } \ \ .
\label{c.19}
\eeqn
The same procedure is used in the evaluation of the eigenvalue density.
In radial coordinates, the integral for the Gaussian density is
\beqn
 \rho_G (\lambda ) =e^{-a\lambda^2}{1\over z_G}\int_0^\infty  dr r^{n^2-2}
e^{-ar^2}\int_{\omega (n-1, 1)} da_{n-1} \Delta^2_{n-1} \prod_{i=1}^{n-1}
({\lambda\over r} -\lambda_i)^2 \ \ .
\label{c.20}
\eeqn
The surface contribution is much alike the one in the expression (\ref{c.17})
for $\rho_{\delta }(\lambda )$. To implement this similarity, we introduce
the expansion
\beqn
\int_{\omega (n-1,1)}da_{n-1}  \Delta^2_{n-1} \prod_{i=1}^{n-1}(x
-\lambda_i)^2 =\sum_{k=0}^{2n-2} c_k x^k \ \ .
\label{c.21}
\eeqn
Since $\rho_G(\lambda )$ is even in $\lambda$, only the even coefficients
are different from zero. The expressions for the densities in the two
ensembles are:
\beqn
\rho_G (\lambda ) =e^{-a\lambda^2}{1\over z_G}a^{-{1\over 2}(n^2-1)}
{1\over 2}\sum_{k=0}^{n-1} c_{2k} (\lambda\sqrt a)^{2k} \Gamma \left (
{{n^2-1}\over 2} -k \right ) \ \ ,
\label{c.22}
\eeqn
\beqn
\rho_{\delta } (\lambda ) =
\frac{n}{ 2 z_{\delta }}(R^2-\lambda^2)^{ \frac{(n-1)^2}{2} -1}
\sum_{k=0}^{n-1} c_{2k} \lambda^{2k}(R^2-\lambda^2)^{n-1-k}
\quad , \quad R^2 \equiv n A^2\ \ .
\label{c.23}
\eeqn
The coefficients $c_{2k}$ are obtained by comparing the polynomial
expression in (\ref{c.22}) and the exactly known result (\ref{c.13}):
\beqn
 c_{2k}= 2^{1-\frac{n^2}{2} } \frac{ (2 \pi)^{\frac{n}{2}} (-4)^k} 
{ \sqrt{ \pi}
\,  (2k)!} \frac{ \prod_{j=1}^n  \,j!}{      
\Gamma \left ( {{n^2-1}\over 2} -k \right )}
\frac{1}{ n}\sum_{\ell=k}^{n-1}(-1)^\ell
{{(2\ell )!}\over {2^\ell \ell ! (\ell-k)!}} \pmatrix {n\cr \ell +1\cr } \ \ .
\label{c.24}
\eeqn
More explicitly, the spectral densities for the lowest order random
matrices are
\beqn
\rd(\la)&=& \frac{1}{\pi \, \sqrt{2 A^2 -\la^2} }
\quad , \quad \mbox{ for} \; \; n=2 \; ,
\nn \\
\rd(\la)&=& \frac{35 \sqrt{3} }{576 A^7} ( A^2 -\frac{\la^2}{3} )
\bigg[  3 A^4-2 \la^2  A^2+ 3 \la^4
\bigg] \; , \; \mbox{ for} \; \; n=3 \; ,
\nn \\
\rd(\la)&=& \frac{32}{429 \pi A^{14}} \bigg( A^2 -\frac{1}{4}\la^2 \bigg)^{
\frac{7}{2}} \bigg[ 12 A^6 +30  \la^2  A^4-53 \la^4  A^2+38  \la^6
\bigg] \; , \; \mbox{ for} \; \; n=4 \; ,
\nn \\
\rd(\la)&=&\frac{2028117\sqrt{5}}{5^4\,(2 A)^{23}} (A^2 -\frac{\la^2}{5} )^7
\bigg[375 A^8-300 \la^2  A^6+4490 \la^4  A^4-5996 \la^6  A^2+
2711 \la^8 \bigg]  , \; \mbox{ for} \; \; n=5.
\nn \\
\label{c.25}
\eeqn
To evaluate the spectral density for the bounded trace ensemble , for finite
$n$ , one may proceed in a similar way as in the Gaussian case, to obtain
\beqn
\rho_{\theta}(\lambda)=\frac{1}{z_{\theta}} \int_0^{ \sqrt{R^2-\lambda^2} }
dr \, r^{n^2-2} \, 
\int_{\omega (n-1, 1)} da_{n-1} \Delta^2_{n-1} \prod_{i=1}^{n-1}
({\lambda\over r} -\lambda_i)^2 
\label{c.26}
\eeqn
and therefore
\beqn
\rho_{\theta}(\lambda)=\frac{1}{z_{\theta}} (R^2-\lambda^2)^{ \frac{n^2-1}{2} }
\sum_{k=0} ^{n-1} \frac{c_{2 k}}{n^2 - 2 k -1} \left ( \frac{\lambda^2}{
R^2-\lambda^2} \right )^k
\quad , \quad R^2 \equiv n A^2 \ \ .
\label{c.27}
\eeqn
The same result may be obtained by inverting the differential
equation (\ref{c.8}).
In a similar way , it is possible to write the explicit expressions
of the two-point correlator of restricted trace ensembles in terms of
the known  two-point correlator of the Gaussian ensemble at finite $n$. 
We obtain
\beqn
\rho_G (\lambda, \mu )={1\over {2z_G}} e^{-a(\lambda^2+\mu^2)}
a^{-{1\over 2}(n^2-4)} \sum_{r,s=0}^{2n-2} c_{r,s}(\lambda\sqrt a)^r
(\mu\sqrt a)^s \Gamma \left ( {{n^2-r-s}\over 2} -1\right ) \ \ ,
\label{c.28}
\eeqn
\beqn
\rho_\delta (\lambda, \mu )={1\over {z_\delta }} (R^2-\lambda^2-\mu^2)^
{ {1\over 2}(n^2-3)}\sum_{r,s=0}^{2n-2} c_{r,s}\lambda^r \mu^s
(R^2-\lambda^2-\mu^2)^{-{{(r+s)}\over 2} } \ \ ,
\label{c.29}
\eeqn
where the coefficients $c_{r,s}=c_{s,r}$ are defined by
\beqn
(x-y)^2\int_{\omega (n-2,1)}da_{n-2}\Delta^2_{n-2} \prod_{k=1}^{n-2}
(x-\lambda_i)^2(y-\lambda_i)^2 = \sum_{r,s=0}^{2n-2} c_{r,s}x^ry^s \ \ .
\label{c.30}
\eeqn

\vspace{1.5cm}

\setcounter{equation}{0}
\section{Generalized restricted trace ensembles at large n}

With some
generality, for an arbitrary polynomial potential $V(M)=\sum g_kM^k$, where
$M$ is an Hermitian $n\times n$ matrix, we define the
 generalized {\bf fixed trace ensemble} and the generalized
 {\bf bounded trace ensemble}
 by the two probability densities :
\beqn
{\cal P}_\delta (M) \equiv {1\over {{\cal Z}_\delta}} \, \delta 
\left ( A^2-{1\over n} \mbox{ Tr}\,V(M)\right ) \ \ ,
\label{f.1}
\eeqn
\beqn
{\cal P}_\theta (M) \equiv {1\over {{\cal Z}_\theta}} \, \theta 
\left ( A^2-{1\over n} \mbox{ Tr}\,V(M)\right ) \ \ ,
\label{f.2}
\eeqn
where ${\cal Z}_\delta $  and ${\cal Z}_\theta $ are the normalization factors,
and we used the same notation of the previous section, where they
 correspond to the simplest case $V(x)=x^2$. \\
Both ensembles are invariant under the action of the unitary group. Therefore,
when changing matrix the parameterization from $n^2$ independent matrix
elements to the $n$ real eigenvalues and the parameters for eigenvectors, the
measures factorize into a part given by the Haar measure of $SU(n)$ and a
part involving only the eigenvalues. The latter provides the joint probability
density of the eigenvalues, the starting point for all spectral statistics.
Letting $\phi $ stand for the $delta $ or the $theta $ function, the expression
for the joint probability density is:
\beqn
P_\phi (\lambda_1,\ldots,\lambda_n) = {1\over {z_\phi}} \phi \left (
A^2-{1\over n}\sum_{i=1}^n V(\lambda_i)\right ) \Delta^2
(\lambda_1, \ldots, \lambda_n) \ \ ,
\label{f.3}
\eeqn
\beqn
z_\phi = \int \prod_{i=1}^n d\lambda_i  \, \phi (A^2-{1\over n}
\sum_{i=1}^n V(\lambda_i) )  \, \Delta^2(\lambda_1, \ldots, \lambda_n)  \ \ .
\label{f.4}
\eeqn
The two ensembles are obviously related by the differential equation
analogous to eq.(\ref{c.8}):
\beqn
{\cal P}_{\delta} (\lambda_1,\ldots, \lambda_n) = \left ( 1+\frac{ z_\theta}{
z_\delta }{\partial\over {\partial A^2}}\right ) 
{\cal P}_\theta (\lambda_1,\ldots , \lambda_n) \ \ ,
\label{f.5}
\eeqn
which will be used to study the properties of the bounded trace ensemble
from a knowledge of the fixed trace one. Indeed, in this general setting, 
the latter is easier to evaluate in the large-$n$ limit.\\
Besides the two restricted trace ensembles, it is useful to consider also
the "canonical" ensemble, with same potential $V(M)$ and a parameter $K$:
\beqn
{\cal P}(\lambda_1,\ldots, \lambda_n) ={1\over {z}}\  
e^{-K n\sum_{i=1}^n V(\lambda_i) } \Delta^2(\lambda_1,\ldots ,\lambda_n) \ \ ,
\label{f.6}
\eeqn
\beqn
z=\int \prod_{i=1}^n d\lambda_i \  
e^{-K n\sum_{i=1}^n V(\lambda_i) } \Delta^2(\lambda_1,\ldots ,\lambda_n) \ \ .
\label{f.7}
\eeqn
As it is well known, the partition function for the eigenvalues may be given
the interpretation as the partition function of a one dimensional gas of $n$
particles with pairwise repulsive interaction and, in the canonical case,
subject to the external potential $V(\lambda )$. In the restricted trace
ensembles the potential enters as a constraint depending on the positions
of {\sl all} particles. This main difference makes the analysis of these
models difficult and interesting, especially for the issue of the universality
properties of correlators.\\
While for "canonical" models the powerful technique of orthogonal polynomials
applies, giving at least formally and for any value of $n$ the explicit
expressions of all correlators, for the restricted trace ensembles we must
content ourselves with the analysis of the eigenvalue density in the large
$n$ limit. This is easily done for the fixed trace ensemble, whose
$\delta $ constraint can be taken into account in the energy
functional through a Lagrange multiplier. In the large-$n$ limit, the
eigenvalue configuration is described by a normalized density
$\rho (\lambda) $, and the energy functional associated to it is
\beqn
H[\rho ] = -\int d\lambda d\mu \rho (\lambda) \rho (\mu ) \log |\lambda -
\mu | +\alpha \left ( A^2-\int d\lambda \rho (\lambda )V(\lambda )\right ) +
\beta \left ( 1-\int d\lambda \rho (\lambda )\right ) \ \ .
\label{f.8}
\eeqn
The saddle point configuration is the one that minimizes the energy, and
is precisely the sought limit density $\rho_\delta $. It solves the
following equation, valid for any $\lambda $ inside the unknown support $L$ of
$\rho_\delta $:
\beqn
 0 = {{\delta H[\rho]}\over {\delta \rho (\lambda )}} =
-2\int d\mu \rho_\delta (\mu ) \log |\lambda -\mu |  -\alpha V(\lambda ) 
-\beta \ \ .
\label{f.9}
\eeqn
A derivative in $\lambda $ eliminates the parameter $\beta $ associated to the
constraint of normalization, and yields a Cauchy-Hilbert integral equation for
the limit density:
\beqn
 \Pint_L d\mu {{\rho_\delta (\mu )}\over {\lambda-\mu }} =
{\alpha \over 2}V^\prime  (\lambda ) \ \ , \quad\quad \lambda\in L \ \ .
\label{f.10}
\eeqn
For any $\alpha $, which is still unknown, and after having fixed a
geometry for the support $L$ (an interval, for example) the equation 
(\ref{f.10}) is
solved using  analyticity arguments, and the extrema of $L$ are fixed by the
normalization condition \cite{brezin}. Inside the family of pairs $L(\alpha)$ 
and
$\rho_\delta(\lambda;\alpha )$ 
parameterized by $\alpha $, the pair that describes the
large-$n$ limit of the fixed trace ensemble is determined by the value
$\alpha=\overline\alpha $, solution of the equation
\beqn
A^2 = \int_{L(\overline \alpha )} d\lambda \rho_\delta
(\lambda ;\overline\alpha ) V(\lambda ) \ \ .
\label{f.11}
\eeqn
The number $\overline\beta $ of the extremal solution may be evaluated from
eq. (\ref{f.9}) by choosing a convenient value of $\lambda $ in $L$.\\
The density $\rho_\delta $ so far obtained, coincides with the limit density
of the canonical model (\ref{f.7}), with parameter $K=\overline\alpha $.
In the particularly simple case $V(M)=M^2$, one obtains also for the restricted
trace ensemble a limit density described by Wigner's semicircle law, with
radius $ 2 A $.
The energy functional (\ref{f.8}) evaluated at the extremum, is
\beqn
H[\rho_\delta]=-\int d\lambda d\mu \rho_\delta (\lambda) \rho_\delta (\mu )
\log |\lambda -\mu | \ \ ,
\label{f.12}
\eeqn
where the double integral may be simplified by using the equation (\ref{f.9}) 
and the constraints:
\beqn
\int d\lambda d\mu \rho_\delta (\lambda) \rho_\delta (\mu )\log
|\lambda -\mu | =-{1\over 2}\overline \alpha A^2 -{1\over 2}\overline\beta\ \ .
\label{f.13}
\eeqn
We then obtain  the large-$n$ expression of the partition function
\beqn
z_\delta \to e^{-{1\over 2} n^2 (\overline \alpha A^2 + \overline\beta )} 
\equiv  e^{-n^2 f(A^2)} \ \ .
\label{f.14}
\eeqn
Since $z_\delta =\frac{\partial}{\partial A^2} z_{\theta}$, eq.(\ref{f.14})
implies
\beqn
\frac{z_{\theta} }{z_\delta}=\frac{ {\cal Z}_{\theta} }{ {\cal Z}_\delta} \to
- \frac{1}{n^2 \frac{\partial}{\partial A^2}  f(A^2) } \ \ .
\label{f.15}
\eeqn
A simple check  is provided by the monomial potentials
$V(x)=x^{2k}$. In this simple case, the normalization
constants $z_{\theta} $ and $z_\delta$ may be evaluated by a rescaling
of the eigenvalues with the result $\frac{z_{\theta} }{z_\delta}=
\frac{2 k A^2}{n^2} $. \\
The  result (\ref{f.15}) is most useful and it implies the generalization
of eq.(\ref{c.8})
\beqn
<O(M)>_{\de }\ = \bigg( 1+ c_n \frac{\partial}{\partial A^2}
 \bigg)
<O(M)>_{\th } \quad ; \quad c_n \to -
 \frac{1}{ n^2 \frac{\partial}{\partial A^2}  f(A^2) } \ \ .
\label{f.16}
\eeqn
By using this equation both for the spectral density and for the
density-density correlator (\ref{i.4}), we obtain an exact equation,
for any $n$ :
\beqn
\rho_{\delta , c} (\lambda, \mu)&=&\rho_{\delta}(\lambda, \mu)-
\rho_{\delta}(\lambda) \rho_{\delta}(\mu)=
\nonumber \\
&=& \bigg( 1+c_n  \frac{\partial}{\partial A^2}\bigg)
\rho_{\theta}(\lambda, \mu)-
 \bigg( 1+ c_n \frac{\partial}{\partial A^2}\bigg)
\rho_{\theta}(\lambda)
 \bigg( 1+c_n \frac{\partial}{\partial A^2}\bigg)
\rho_{\theta}(\mu)=
\nonumber \\
&=& \rho_{\theta , c}(\lambda, \mu) +
 c_n \frac{\partial}{\partial A^2}
 \rho_{\theta , c}(\lambda, \mu) -
  (c_n)^2 \bigg(\frac{\partial}{\partial A^2}\rho_{\theta}(\lambda)
\bigg) \bigg(\frac{\partial}{\partial A^2}\rho_{\theta}(\mu)\bigg) \ \ .
\label{f.17}
\eeqn
We have not proven that the generalized restricted trace matrix ensembles
have a topological expansion in the "large-$n$" limit and the factorization
of invariant operators, analogous to matrix ensembles defined by
"canonical" probability densities. The analysis of next section,
where the fixed trace constraint is reached as a limit of  the probability
density ${\cal P}_l(M)$ indicates that such properties are very
likely. Therefore it seems reasonable to assume, as for the "canonical" 
probability densities,
\beqn
 \rho_{\phi}(\lambda, \mu) \to  \rho_{\phi}(\lambda)
 \rho_{\phi}(\mu) + \frac{1}{n^2} {\overline \rho}_{\phi}(\lambda, \mu)+
O(\frac{1}{n^3}) \ \ ,
\label{f.18}
\eeqn
where $\phi$ stands for the $\delta$ or the $\theta$ functions. This assumption
(\ref{f.18}), as well as more general assumptions, together with
eq.(\ref{f.17}) and eq.(\ref{f.16}) imply , in the "large-$n$" limit
$ \rho_{\delta , c}(\lambda, \mu)= \rho_{\theta , c}(\lambda, \mu)$.
The results of this section are rather general and formal. The determination
of the Lagrange multiplier $\overline\alpha$ in eq.(\ref{f.11})  of course
depends on the model potential $V(M)$ in a non trivial way and on the
various phases of the model. We provide a specific example in the next
section, by the study of the  potential $V(M)=g_2M^2+g_4 M^4$.

\vspace{1.5cm}

\setcounter{equation}{0}
\section{Phase transitions}

In the previous section it was shown that, in the large-$n$ limit, the spectral
density of restricted trace ensembles with polynomial potential $V(\lambda, 
g_i)$, where $g_i$ are the couplings, coincides with the spectral density
of the "canonical" ensemble with potential ${\bar\alpha }V(\lambda, g_i)$.
The scaling factor ${\bar\alpha }$, solution of eq. (\ref{f.11}), is actually
a nonlinear function of the couplings $g_i$. The correspondence between the
two sets of parameters, namely $g_i$ and ${\bar \alpha}g_i$, is one-to-one
only in the perturbative phase.\\
In this section we show in detail the case of the even quartic potential 
\beqn
V(M)  = g_2 \, M^2  + g_4 \, M^4  \ \ .
\label{d.2}
\eeqn
where the nonlinear relation originates different phase diagrams. To this
end, we find it useful
to consider the {\bf squared trace ensemble} $\Pl (M)$
\beqn
\Pl (M) &=& \frac{1}{\Zl}\exp\left[ -l\left(-2 n A^2\Tr V(M) 
\ +\ (\Tr V(M))^2\right)\right] \ \ ,\nn \\
\Zl &=& \int\dM \exp\left[ -l\left(-2n A^2\Tr V(M) 
\ +\ (\Tr V(M))^2\right)\right] \ \ . \l{Pcross}
\label{d.1}
\eeqn
The large-$n$ limit of the model described by the
probability distribution $\Pl (M)$ is easily found by
the saddle point approximation. These type of models,
where the exponent of the Boltzmann weight is a  sum
of different powers of traces of even powers of the
random matrix  was 
analyzed in several matrix models in zero and one dimension
\cite{das} - \cite{david}.
The additional "trace-squared" terms  were interpreted 
to provide  touching interactions to the dynamical triangulated
surfaces defined  by the matrix potential $ {\Tr}\, V(M)$. \\
For any fixed $l$, the model in eq.(\ref{d.1})-(\ref{d.2})
is equivalent in the large-$n$ limit
to a  random matrix ensemble with the well studied
"canonical"  probability distribution
\beqn
\Pc (M) &=& \frac{1}{\Z}\exp(-n\Tr V(M)) \ \ ,\nn \\
\Z     &=& \int\dM \exp(-n\Tr V(M)) \ \ ,\ \ 
V(M)  = g_2' \, M^2  + g_4' \, M^4  \ \ ,
\label{d.3}
\eeqn
provided the parameters $g_2'$ and $g_4'$ are suitable functions of the
parameters of  the model in eqs.(\ref{d.1})-(\ref{d.2}). This may be
accomplished by two equations, such as the requirement
that the expectations of $<{\Tr} \, M^2>$ and  $<{\Tr} \, M^4>$
should be the same for the two probability distributions.\\
On the other hand, for fixed $n$, in the large-$l$ limit,
 $\Pl (M)$ reproduces precisely the generalization of the fixed  trace ensemble
 $\Pd(M)$ , as one sees from the following
representation of the $\de$-function
$\de (x)=\lim_{l \to \infty}\sqrt{\frac{l}{\pi}}\exp(-lx^2)$. Of course,
when choosing $g_4=0$, we merely reobtain the results of the analysis by 
Bronk and Rosenzweig.\\
Let us now recall the  saddle point
analysis for the large-$n$ limit of the ensemble $\Pl (M)$,
eqs.(\ref{d.1})-(\ref{d.2}). Since it proceeds along well known
analysis, we include, for more generality the cases of the random
matrix $M$ belonging to the orthogonal, unitary, or symplectic
ensembles, corresponding to the parameter $\beta=1, 2$ or $4$.
It is important to notice that, unlike the familiar quartic
probability distribution (\ref{d.3}), the probability distribution
(\ref{d.1})-(\ref{d.2}) is well defined for any real value of the
two parameters $g_2$, $g_4$. Let us begin by assuming $g_2>0$, $g_4>0$,
which corresponds to the perturbative (or one-cut)
phase ; later in the section  the complete phase diagram will be
described.
For any finite  positive value of the parameter
$l$,  the density of eigenvalues $\rho_l(\la)$ is the solution
of the singular integral equation
\beqn
 \beta ~\Pint  d\mu \frac{\rho_l(\mu)}{\la -\mu} \ = \ 
2l\left(g_2c_2+g_4c_4-A^2\right) 
\, V' (\la) \ = \  2 g_2'  \la~+~ 4 g_4'   \la^3
\label{d.4}
\eeqn
where the moments $c_k $  are defined by
\beqn
c_k\ \equiv\ \int d\la \, \la^k \rl(\la)  
\label{d.5}
\eeqn
and $g_k'$ are the effective couplings : 
\beq
g_k'\ =\ 2l\left(g_2c_2+g_4c_4-A^2\right) g_k \ \ .
\l{geff}
\eeq
From the symmetry of the potential the support of  $\rho_l(\la)$
is expected to be one segment or two segments, in either case
symmetric with respect to the origin.
The solution of the saddle-point equation (\ref{d.4}) in the one segment 
phase reads
\beqn
\rho_l(\la) \ =\ \frac{2}{\beta \pi}(g_2'~+~g_4'b^2~+~ 2g_4'\la^2)
                  \sqrt{b^2-\la^2} \ \ ,
\label{d.6}
\eeqn
where the endpoint of the support $[-b,b]$ 
is given by the normalization condition on the eigenvalue density
\beqn
1\ =\ \int_{-b}^{b}d\la \rho_l(\la) \ =\ 
2l\left(g_2c_2+g_4c_4-A^2\right) 
\frac{b^2}{2 \beta}(2g_2 + 3b^2g_4) \ \ ,
\label{d.7}
\eeqn
where we have used again the $g_k$'s. \\
The moments $c_2$ and $c_4$ can be obtained when requiring self consistency
by inserting the solution eq. (\ref{d.6}) back into the definitions
(\ref{d.5}), which yields the linear system of equations
\beqn
c_2 &=&\frac{2}{\beta}\, 2l\left(g_2c_2+g_4c_4-A^2\right) 
\frac{b^4}{8}(g_2 + 2b^2g_4) \  \ ,\nn \\
c_4 &=&\frac{2}{\beta}\, 2l\left(g_2c_2+g_4c_4-A^2\right) 
 \frac{b^6}{64}(4g_2 + 9b^2g_4)   \ \ . 
\label{d.8}
\eeqn
For a potential of higher degree we will again get a linear system of
equations for the corresponding moments $c_k$, $k\!=\!1,..,m$, which is due 
to the fact that the solution of the saddle-point equation 
will again depend linearly on the coupling constants $g_k'$ as in eq. 
(\ref{d.6}). Instead of solving the eqs. (\ref{d.8}) for $c_2$ and $c_4$
we can also express them entirely in terms of the couplings with the help of
eq. (\ref{d.7})
\beqn
c_2 &=& \frac{b^2(g_2 + 2b^2g_4)}{2(2g_2+3b^2g_4)} \ \ ,\nn \\
c_4 &=&  \frac{b^4(4g_2 + 9b^2g_4)}{16(2g_2+3b^2g_4)}   \ \ . 
\label{d.9}
\eeqn
The same trick can be used to express the eigenvalue density
eq. (\ref{d.6}) only in terms of the $g_k$, which reads
\beqn
\rho_l(\la) \ =\ \frac{4}{\pi b^2(2g_2+3b^2g_4)}
      \left(g_2~+~g_4b^2~+~ 2g_4\la^2\right) \sqrt{b^2-\la^2} \ \ .
\label{d.10}
\eeqn
where the endpoint of the support $b$ is the root of the fourth order 
equation in $b^2$
\beqn
l \bigg(9 (g_4)^2 b^8 ~+~ 20 g_4 g_2 b^6 ~+~ 8( (g_2)^2-6 A^2 g_4 ) b^4 ~-~
32 A^2 g_2  b^2 \bigg) =16 \beta
\label{d.11}
\eeqn
which, for vanishing $g_4$ and  positive $g_  2$ is asymptotic
to $ b^2 \sim  2(A^2+\sqrt{A^4+\beta/(2l)})/g_2$.
By comparing eq(\ref{d.4}) with the analogous saddle point equation 
for the "canonical" 
quartic probability distribution (\ref{d.3}) it is obvious
that they have the same eigenvalue density,
in the large-$n$ limit, for both phases of the
model, provided the effective coupling eq. (\r{geff}) are precisely identified
with those of the "canonical" distribution
\beqn
g_2' &=& 2l\left(g_2c_2+g_4c_4 -A^2\right)g_2 =
\frac{2 \, \beta}{b^2(2  g_2+3b^2  g_4)}  g_2\ \ ,
\nn \\
g_4'&=& 2l\left(g_2c_2+g_4c_4 -A^2\right)g_4=
\frac{2 \, \beta}{b^2(2  g_2+3b^2  g_4)} g_4\ \ .
\nn \\
\label{d.12}
\eeqn
Of course, the last equality on the right sides of previous equations only
holds in the one cut phase.
For simplicity, let us now proceed with $\beta=2$.
In terms of $g_2'$ and $g_4'$, the equation for the support (\ref{d.11}) is the
more familiar equation $3 g_4' b^4+2 g_2'  b^2-4=0$. The phase diagram
of the "canonical" quartic model $\Pc (M)$, eq.(\ref{d.3}), is well known:
if  $g_2'$ is fixed positive ,
the one-cut solution (\ref{d.6})-(\ref{d.7}) holds for any real
$g_4'$ such that 
\beqn
g_4' \geq  -\frac{1}{12} (g_2')^2 \ \ ,
\label{d.13}
\eeqn
which is a border of existence for the model. If $g_4'$ is fixed positive,
the one-cut solution  holds for any real $g_2'$ such that 
\beqn
g_2'  \geq   -2 \sqrt{g_4'} \ \ ,
\label{d.14}
\eeqn
which is the line of phase transition to the symmetric two-cut solution :
\beqn
\rho_l(\la) \, =\, \frac{2 g_4' | \lambda| }{\pi}
                  \sqrt{(D^2-\la^2)(\la^2-C^2)} \ \ ,
\label{d.15}
\eeqn
with ends of support $[-D \,,\,-C\,]\ \cup\ [\,C \, , \, D\,]$ being 
solutions of
\beqn
g_2' +g_4' (C^2+D^2)=0 \quad , \quad
g_4' (D^2-C^2)^2=4 \ \ 
\label{d.16}
\eeqn
The map between $\{g_2 , g_4 \}$ and  $\{g'_2 , g'_4 \}$ in this phase,
may be found after the evaluation of $ \{c_2 , c_4 \}$ and the
requirement of self-consistency , just as before. \\ 
It is straightforward to see that the phase transition line (\ref{d.14}) 
becomes, in the parameters of the model (\ref{d.1})-(\ref{d.2}), the
couple of lines
\beqn
g_4= \frac{l}{4}\, (g_2)^2  \bigg(-A^2 \pm \sqrt{A^4-3/(2l)} \bigg)
\quad , \quad g_2>0 \quad , \quad g_4<0 \ \ .
\label{d.17}
\eeqn
Therefore if $A^4-3/(2l)<0$ the model (\ref{d.1})-(\ref{d.2}) has the
one-cut solution for every real value of $g_2 $ , $g_4$. In the other
case $A^4-3/(2l)>0$ the two-cut solution holds in the region
of parameters bounded by the two curves $(\ref{d.17})$, while the 
one-cut solution holds everywhere else in the plane of
real values of $g_2 $ , $g_4$.\\
The image of the existence line (\ref{d.13}) , in the space of parameters
$g_2 $ , $g_4$ is a couple of curves :
\beqn
g_4 &=& \frac{l}{12}\, (g_2)^2 \big( A^2-\sqrt{A^4+7/(6l)} \big)
\quad , \quad g_2>0 \quad , \quad g_4<0 \; ,
\nonumber \\
g_4 &=& \frac{l}{12}\, (g_2)^2 \big( A^2+\sqrt{A^4+7/(6l)} \big)
\quad , \quad g_2<0 \quad , \quad g_4>0 \; .
\label{d.18}
\eeqn
There are two  regions of the plane of real variables $g_2 $ , $g_4$ :
the first one
bounded by the first line (\ref{d.18}) and the positive axis $g_2$,
and the second one bounded by the second line (\ref{d.18})
and the negative axis $g_2$, where the equation of the support
(\ref{d.11}) of the one-cut solution has three possible values.
The one-cut solution (\ref{d.10}) as function of the parameters
$g_2 $ , $g_4$ has a first order discontinuity in these regions
due to the cubic type instability of the solution of the eq.(\ref{d.11})
with respect to the parameters $g_2 $ , $g_4$ . As usual, the
lines of the first order transition are determined by comparing the 
evaluation of the free energy of the model, as functions of the different
possible values of the endpoint of the support $b$. \\
In the remaining part of this section, we consider 
 the limit $l\to\infty$ where we obtain the distribution
$\Pd(M)$ with the potential (\r{d.2}) explicitly. We shall denote
$\overline{c}_k\!\equiv\!\lim_{l\to\infty}c_k$ and 
$\overline{b}\!=\!\lim_{l\to\infty}b$ .
Because of the $\de$-function in the distribution it will hold
\beq
A^2\ =\ g_2\overline{c}_2~+~g_4\overline{c}_4 \l{Rc}
\eeq
 whereas the quantity
$l(g_2c_2+g_4c_4-A^2)$ will stay finite, as one can see from
eq. (\r{d.7}). Eq. (4.20) is actually eq. (\ref{f.11}) for the quartic
potential considered in this section.  Eq. (\ref{d.4}) shows that the 
model with $\Pd(M)$ has the same
eigenvalue density of the "canonical" quartic model (\ref{d.3}),
provided $g_2'=\overline\al g_2$ , and $g_4'=\overline\al g_4$, where
\beqn
\overline\al &=& 
\lim_{l\to\infty} 2\, l~(g_2c_2+g_4c_4-A^2) \ = \nn\\
 &=& \left(\frac{ \overline{b}^2}{4}(2g_2 + 3 \overline{b}^2 g_4)\right)^{-1} 
  \ \ . \l{Aexpl} 
\eeqn
The results for the moments eqs. (\r{d.9}) and the density 
eq. (\r{d.10}) carry over when replacing everything 
by barred quantities. Eq. (\ref{Aexpl}) gives the solution
to eq. (\ref{f.11}) and shows its dependence
on the coupling constants of the quartic potential eq. (\ref{d.2}). \\
The phase diagram for $l=\infty$ is similar to the  one
previously described for finite $l$, with some simplifications.
The couple of lines (\ref{d.17}) which are boundaries of the
two-cut phase become the line
\beqn
g_4=-\frac{3}{16}\, \frac{(g_2)^2}{A^2} 
\quad , \quad g_2>0 \quad , \quad g_4<0 \ \ ,
\label{d.20}
\eeqn
and the negative $ g_4$ axis. Similarly there are two regions of
multiple solution for the one-cut support, where a first order
discontinuity will occur. One is bounded   by the positive part
of the $g_2$ axis and the line
\beqn
g_4=-\frac{7}{144}\, \frac{(g_2)^2}{A^2} 
\quad , \quad g_2>0 \quad , \quad g_4<0 \ \ .
\label{d.21}
\eeqn
The second region is the entire region $g_4>0$ and $g_2<0$.
Eq.(\ref{Rc}) for the endpoint $\overline{b}$ of the one-cut
solution turns into
\beqn
0\ =\ \overline{b}^2  \bigg[
9(g_4)^2\overline{b}^6 ~+~ 20g_2g_4\overline{b}^4 ~+~ 8((g_2)^2-6g_4A^2)
\overline{b}^2 ~-~ 32g_2A^2   \bigg]    \ \ .
\label{d.22}
\eeqn
The vanishing support $\overline{b}=0$ actually provides the limiting 
solution $\rd(\la)= \delta(\lambda)$ in the sector $g_2<0$, $g_4<0$.
In other regions of parameter space the support is determined by
the solution of the third order equation in $\overline{b}^2 $ above.\\
Let us finally extract the result for the Gaussian distribution
${\cal P}_{\delta}(M)$ 
with potential $V(M)\!=\!g_2M^2$ from the above formula 
by setting $g_4\!=\!0$. Eq. (\ref{d.22}) leads to 
\beq
\overline{b}^2 \ =\ \frac{4A^2}{g_2}
\eeq
with the corresponding eigenvalue density from eq. (\r{d.10})
\beq
\rho_{\delta}(\la)
 =\ \frac{2}{\pi \overline{b}^2}\sqrt{\overline{b}^2-\la^2} \ \ .
\eeq
this is the well known semi-circle spectral density and
together with eq.(\ref{c.8} ) it reproduces the old result 
 \cite{rosenzweig} that the spectral density of the restricted trace ensembles
is equal, in the "large-$n$" limit, to the 
spectral density of the Gaussian ensemble.

\vspace{1.5cm}
\noi
{\sc Acknowledgment:}\\
The work of G.A. is supported by European Community grant no. ERBFMBICT960997.
It particular he wishes to thank the Physics Department of Parma for its
warm hospitality while part of this work was being done.

\end{document}